\documentclass[aps,prl,twocolumn,superscriptaddress,showpacs]{revtex4-1}
\usepackage{graphicx}

\begin{document}

\title{Deflection of proton beams by crystal miscut surface}

\author{A.A. Babaev}
\email[]{krass58ad@mail.ru}
\affiliation{INFN Sezione di Roma, Piazzale Aldo Moro 2, 00185 Rome, Italy} \affiliation{Tomsk Polytechnic University, Lenin Ave
40, 634050 Tomsk, Russia}
\author{G. Cavoto}
\affiliation{INFN Sezione di Roma, Piazzale Aldo Moro 2, 00185 Rome, Italy}
\author{S.B. Dabagov}
\affiliation{INFN Laboratori Nazionali di Frascati, Via E. Fermi
40, 00044 Frascati, Italy} \affiliation{RAS P.N. Lebedev Physical
Institute \& NRNU MEPhI, Moscow, Russia}

\date{\today}

\begin{abstract}
First computer experiment results on proton beam deflection by the crystal miscut surface are presented. The phenomenology of proton channeling and quasichanneling has been applied to describe new features of the beam deflection. The analysis predicts efficient beam deflection by the acute crystal end due to repelling miscut potential.
\end{abstract}


\maketitle


One of the most important problems in accelerator physics is known as a beam collimation problem, which includes  the beam shaping, the halo removing, etc. Usually to improve the beam quality the complex system of magnets as well as solid collimators and beam scrapers are used. The methods based on bent crystal technology open new ways to efficient control the proton or heavy ion beams especially at ultra-high energies \cite{Tsyg1,Murp1}. The applicability of bent crystal collimators is based on the effective particle deflection in the crystal volume due to the planar channeling \cite{Scan1,Scan2} or volume reflection \cite{Scan3,Scan4} phenomena. In general, both effects take place when fast charged particles penetrate into the crystal at small glancing angle $\theta_0$ to crystallographic planes. In this case the motion of a fast projectile is characterized as a quasi free longitudinal motion inside the crystal being transversally trapped by so-called averaged planar potential \cite{Lind1,Gemm1}.

 For positively charged particles the averaged potential of a single plane has the maximum at the plane position, decreases rapidly at the increase of the distance from it, and, finally, the electric field becomes negligible at the inter planar distance. Thus, two neighboring planes form the potential well with the minimum between them.

 The particle channeling takes place if $\theta_0<\theta_L$, where at the particle ultra-relativistic energies $E$ the critical angle $\theta_L=(2U_0/E)^{1/2}$, known as Lindhard angle, with the deep value $U_0$ of a planar potential. The planar channeled particle is trapped by the potential well between two adjacent planes and moves along these planes oscillating between them. If the incident angle $\theta_0$ is of the order of a few critical angles $\theta_L$, the particle motion is still governed by the averaged potential, but the trajectory is not bound within a single planar channel. This regime of motion is called quasichanneling. The volume reflection appears for a quasichanneled particle at specific conditions \cite{Tara1,Baba1}.

The demands of fine-tuned beam management require very accurate manipulations for a crystal positioning into the beam. In particular, the crystal has to be inserted into the beam at very small distance providing very small impact parameters (i.e. the distances from the crystal edge along the crystal entrance face to the point where the particles hit the crystal). For example, in experiments \cite{Scan1} the averaged impact parameter was estimated to be of the order of 100 nm. Hence, the particles hitting the lateral crystal surface instead of expected front surface can essentially influence the crystal merits to deflect the beam. The lateral surface considered here is directed along the beam propagation making small grazing angle. Thus, moving the crystal into the beam the particles first interact with the lateral surface that makes important studying the features of beam interaction with the crystal surface at small glancing angles.

In principle, the crystals available for modern experiments have almost perfect flat surfaces \cite{Bari1,Bari2} characterized by the ordered atoms location. The angular asymmetry of beam scattering by crystal surface as well as the periodicity in the energy loss spectra of scattered beam were discovered almost together with the channeling effect and witness to the ordered surface lattice \cite{Mash1,Kley1,Sizm1}. The surface channeling was discussed in \cite{Sizm1} when the particle can be first captured into the channeling motion at a surface layer and successfully leave the crystal through the same lateral surface. Moreover, as shown in \cite{Dana1,Dana2,Dana3,Boyk1} the averaged field approximation is valid for description of particle small angle scattering by a crystal surface, which coincides with one of the main crystallographic planes. Recent experiments on surface both scattering and channeling were mostly carried out for non-relativistic light ions. The beam interaction with crystal atomic chains (axial effects) \cite{Nieh1,Have1,Meye1,Robi1,Heil1,Robi2} as well as planes (planar effects) \cite{Boyk1,Heil1,Robi2,Hase1} has been carefully analyzed within various collaborations.

This letter is devoted to studying ultra relativistic particles scattering by lateral crystal surface. The scheme of scattering has been suggested by the geometry of beam collimation experiments. In simulations the beam is oriented at small angle (or parallel) to crystallographic planes to satisfy the planar channeling conditions but at large enough angles to the main crystallographic axes to avoid the axial channeling (see, for example, in \cite{Bari3,Scan8}). The possibility of beam deflection by the  miscut surface is demonstrated. The simulations below are based on the numerical evaluations of particle trajectories in the field of miscut surface. The detailed theory of the effect will be published elsewhere.

\begin{figure*}
 \includegraphics{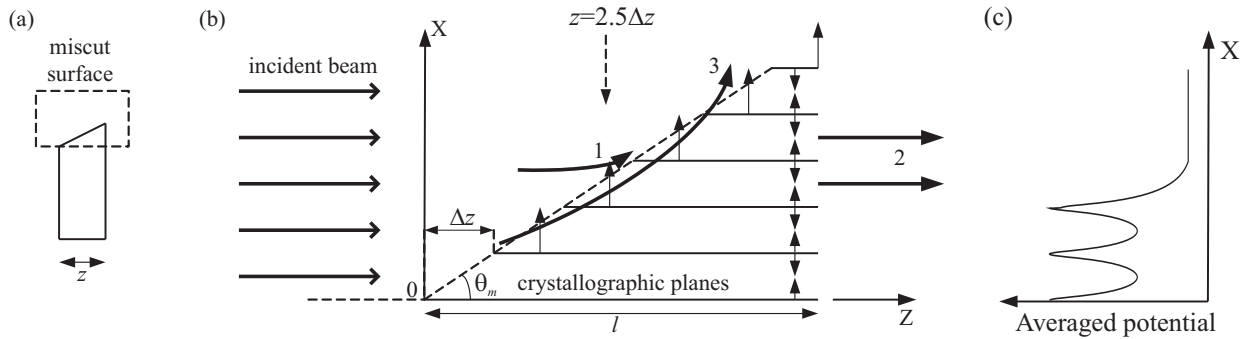}%
 \caption{\label{fig1} (a) Scheme of the crystal with a miscut surface. The crystal has the trapezium shape; the miscut surface is the top triangle part of the trapezium. (b) Scheme of beam deflection by the miscut surface. The crystal field affecting the particle is either the deflecting field at the beginning section $\Delta{}z$ or is the planar channel field (the direction of the field is pointed out by thin vertical arrows). The regimes of a particle motion are pointed out by the bold arrows: 1 --- the particle deflected outward the miscut surface by the repelling potential of the terraces; 2 --- the channeled particles penetrating into the surface layer and successfully leaving the crystal through the back end; 3 ---  the particles deflected via quasichanneling. (c) Scheme of averaged potential $U_{\mathrm{cr}}(x)$ at the longitudinal coordinate $z=2.5\Delta{}z$ pointed out by dash arrow in the panel (b).}
 \end{figure*}

Crystals used in crystal collimation experiments are usually characterized by the lateral surface not parallel to crystallographic planes responsible for the particle channeling in a crystal bulk. The angle between the lateral surface and mentioned planes is known as a miscut angle \cite{Bari1,Else1,Douc1}, while the surface has been called a miscut surface (Fig.~\ref{fig1}(a)). This surface is structured by a set of parallel planes that form stepped terraces; each terrace of the length $\Delta{}z$ is a part of the crystal plane, while a step equals to the inter planar distance (Fig.~\ref{fig1}(b)). Obviously, the miscut angle $\theta_m$ defined by such geometry is extremely small. Positively charged beam hitting the miscut surface of aligned crystal will undergo multiple terrace reflection. Computer simulation results on the repelling action of the miscut surface potential to the beam of relativistic protons will be presented below.

Let consider fast particles of non-divergent beam hitting the crystal surface along the crystallographic planes, as shown in Fig.~\ref{fig1}. If the entry surface has no miscut (the planes are perpendicular to the surface), the particles entering the crystal at channeling conditions become trapped between the planes that define the potential well. On the contrary, in the case of nonzero miscut angle the particles hitting the miscut surface first interact with averaged potential of a single plane that defines the terrace. In our case, protons are reflected by the repelling plane field (terrace potential) outward of the plane. Obviously, we can define three different regimes of proton motion. The first one corresponds to the case when a proton is strongly deflected providing its interaction with next plane (surface potential of upper terrace) instead of averaged channel potential (bulk potential) of adjacent planes, which form lower (actual) and upper terraces (Fig.~\ref{fig1}(b), arrow 1). Both second and third regimes take place when the deflection of a proton is not enough to be out of corresponding planar channel. If herein the angle of deflection, which defines the incident angle, is less than the critical angle of channeling, $\theta_0<\theta_L$ (Fig.~\ref{fig1}(b), arrow 2), the proton will move in a crystal volume being planar channeled. If, however, the angle of deflection at the entrance into the crystal channel exceeds the critical angle of channeling, $\theta_0>\theta_L$, the proton becomes quasichanneled and will suffer a kind of volume reflection (Fig.~\ref{fig1}(b), arrow 3). The kind of motion depends on the initial particle distance to the nearest lower plane at the moment when the particle starts interacting with the crystal field. In general case, all described regimes exist when the beam hits the miscut surface.

The letter aims in studying the particle motion in the miscut surface layer, i.e. in the crystal volume where the plane lengths determined by the miscut angle successfully decrease. Herein, the crystal thickness $l$ corresponds to the maximal plane length. The number of planes $N$ contained in the miscut layer as well as the number of channels $(N-1)$ at fixed $l$ is defined by both miscut angle $\theta_m$ and interplanar distance $a$. The plane section responsible for the repelling field, above named as "a terrace", has the length $\Delta{}z=a/\tan{}\theta_m$. In order to simplify the analysis we should take into account the motion of only protons hitting the "miscut surface", thus, the beam width considered is $(N-1)a$.

  Following the channeling theory the evolution of proton transverse coordinate $x$ is defined by the equation
\begin{equation}\label{eq1}
    \gamma{}m\frac{d^2x}{dt^2}=-\frac{dU_{\mathrm{cr}}}{dx}\,,
 \end{equation}
 where $U_{\mathrm{cr}}$ is the proton potential energy in averaged crystal field, which only depends on the transverse coordinate, $\gamma$ and $m$ are the proton relativistic factor and its rest mass, correspondingly. The averaged planar potential was approximated by Moliere potential \cite{Gemm1}.  The projectile moves along the planes with quasiconstant velocity $v_0$ that defines the longitudinal coordinate of its trajectory  as  a function of time $z(t)=v_0t$. The deviation of projectile trajectory from that  defined by Eq.(\ref{eq1}) due to the proton multiple scattering is taken into account by the technique previously described in \cite{Baba1,Baba2}. The angular distribution of the beam at any fixed moment $\tau$ with respect to the crystal plane (in our case, the (110) plane) is determined by the expressions $\tan\theta_i=v_{xi}(\tau)/v_0$ where $v_{xi}(\tau)=(dx/dt)_\tau$ is the transverse velocity of i-th proton at that moment. This is also valid for the moment, at which the projectile leaves the crystal through either back side or lateral miscut surface. Obviously, as above described, the first corresponds mainly to the channeled protons, while the second - to both quasichanneled in a bulk and deflected by the terrace potential protons.

In our simulations we have used non-divergent 400 GeV proton beam interacting with aligned Si (110) crystal that are typical objects of recent CERN crystal collimation experiments based on the technique of beam channeling in a bent crystal \cite{Scan3,Scan4,Scan5}. The initial beam direction was chosen to be coincided  with the $Z$-axis, hence, the crystal field, which is mostly perpendicular to the (110) planes, has became the function of transverse coordinate $x$. The motion in $Y$-direction, which is directed along the (110) planes and transverse to the initial beam direction, is out of interest in this study and herein has not been considered. Having chosen the initial beam direction along (110) planes of aligned Si crystal, the deflection angle after proton passage through the crystal has been defined  as the angle of a proton velocity with respect to the (110) plane.

\begin{figure}
 \includegraphics{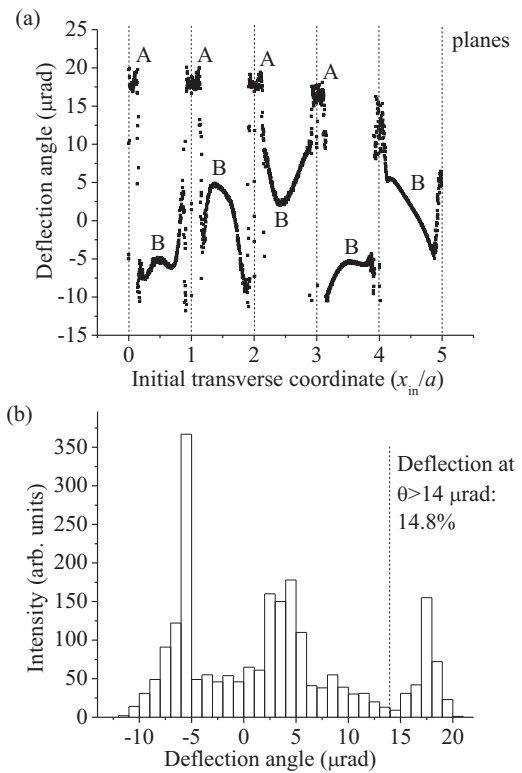}%
 \caption{\label{fig2}Scattering of initially non-divergent beam by miscut surface. (a) Dependence of the deflection angle on the initial transverse particle position, i.e. on the point of penetration into the crystal: A --- the deflected protons, B --- the channeled protons; positions of crystallographic planes composing the miscut surface are pointed out by vertical dashed lines. (b) Angular distribution of protons scattered by miscut surface: the right peak ($\theta>14$ $\mu$rad) corresponds to the miscut deflected protons, while the left side of distribution ($\theta<14$ $\mu$rad) - mainly to the channeled protons.}
 \end{figure}

 The results of simulations for a final deflection angle in dependence on the initial transverse coordinate $x_{\mathrm{in}}$ are presented in Fig.~\ref{fig2}(a). The transverse coordinate is counted from the plane with a maximal length, i.e. from the lowest plane in Fig.~\ref{fig1}. The simulations were performed for the crystal thickness of $l=100\,\mu$m and for the miscut angle of $\theta_{m}=10\,\mu$rad. In this case the miscut layer consists of $N=6$ planes, and each terrace length equals to $\Delta{}z=19$ $\mu$m. Let underline that at the distance of 1 $\mu$m about $10^4$ Si atoms are placed in the crystal plane. Thus, the terrace potential can be described by continuous averaged potential. Also, due to the fact that the terrace length $\Delta{}z\gg{}a$, the field at the distances $0<x<a$ from the terrace plane can be considered to be orthogonal to the plane.

Analysis of the angular distribution shown in Fig.~\ref{fig2}(a) proves the existence of not only proton channeling but the appearance of a separate group of particles at deflection angles $\theta>14$ $\mu$rad. This part of the beam consists of the protons that are strongly deflected outward the miscut surface by the terraces' repelling field, in other words, the protons are reflected by the miscut end of the crystal.

These particles penetrate into the miscut surface near the terrace planes (practically being out of the crystal bulk), where the field is strongest. Hence, they get the significant angular kick, due to which the part of initial beam hitting the miscut surface is essentially deflected ("multiple terrace reflection"). The corresponding angular distribution shown in Fig.~\ref{fig2}(b) clearly reveals the sharp peak of deflected protons. In our specific case we have got the deflected protons intensity of about 15\% from whole beam scattered by the miscut surface. The deflection at such large angles, $\theta\sim17.5$ $\mu$rad, for 400 GeV protons can not be explained by multiple scattering. Indeed, the estimation of multiple scattering for a 400~GeV proton, moving close the plane where the scattering is maximal (see in \cite{Baba1}), results in $5.7\,\mu$rad for the root-mean-square value of deflection angle at $l=100\,\mu$m crystal thickness.

In Fig.~\ref{fig2}(a) one can see, the deflection angle of group~A is maximal for particles penetrating into the crystal near the bottom planes of miscut surface (as depicted in Fig.~\ref{fig1}). As the matter of fact, these protons are mainly deflected through the quasichanneling, i.e their motion is mostly of the kind 3 and not of the kind 1. When the proton moving along the terrace reaches the crystal channel at the incident angle $\theta_0>\theta_L$, this proton will not be bound within that interplane channel. And, thus, it can cross the channel border at the initial $\Delta{}z$ section of the channel, which is formed by next upper terrace of the miscut layer. In this case the proton moves in deflecting single-plane field (the bold arrow 3 in Fig.~\ref{fig1}) getting additional deflection angle $\delta\theta$. After that proton reaches next channel having grown  incident angle $\theta_0+\delta\theta>\theta_0>\theta_L$, and so on. Hence, proton suffers multi plane deflection being quasichanneled that finally results in splitting of the beam into channeled and deflected ones. Obviously, the angle of deflection is proportional to the number of planes crossed.

However, only protons hitting the crystal at the bottom miscut planes enable several planes crossing. Indeed, protons penetrating into the crystal at the top miscut planes do not suffer multi plane interaction due to the limited miscut layer, and are characterized by rather small-angle deflection. These protons as well as protons deflected by upper single planes (the regime 1 in Fig.~\ref{fig1}) fill the interval $10\div14\,\mu$rad between right deflected peak and channeled protons spectrum in Fig.~\ref{fig2}(b).

The crystals used in channeling physics are usually characterized by a miscut angle \cite{Else1}. Nevertheless, the kind of beam deflection described here has not been ever observed in channeling-related experiments. The features of beam scattering by the crystal miscut were mostly examined to prevent negative influence of the miscut to the efficiency of beam collimation based on beam halo deflection by bent crystal planes. Another point is that in performed experiments the beams enter into the crystal mainly through its front surface rather than the lateral miscut surface. Thus, the contribution of miscut deflected part becomes negligible into the total angular distribution. However, its contribution could be essential in the case of nanosize, in cross section, beams.

\begin{figure}
 \includegraphics{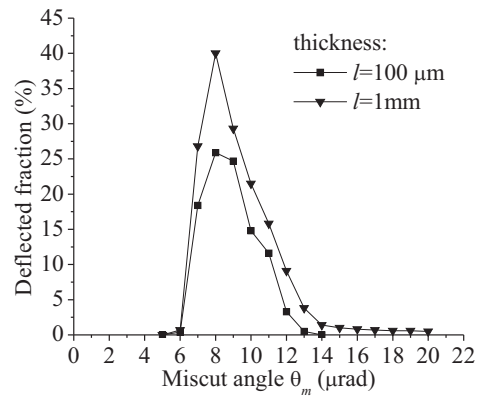}%
 \caption{\label{fig3}Fraction of protons deflected at the angles $\theta>14\,\mu$rad vs miscut angle.}
 \end{figure}

Fig.~\ref{fig3} demonstrates the deflection efficiency as a function of the miscut angle for two crystal thicknesses: $l=100\,\mu$m corresponds to the simulations pointed out in Fig.~\ref{fig2}, while $l=1$ mm represents typical crystal thickness used in bent crystal collimators (see, for example, in \cite{Scan6} and references therein). As seen, the efficient deflection could be observed in very narrow interval of the miscut angles. The deflection efficiency falls rapidly down at the miscut angle increase, which results in decreasing the repelling field length $\Delta{}z$. Simultaneously, increasing the miscut angle gets growing possibility for projectiles to be trapped by the crystal bulk potentials (planar potentials); obviously, this particles will be not reflected by the miscut surface. The suitable miscut angles are very small, $\theta_m\sim10\,\mu$rad, and much less than the typical miscut angles $\sim100\,\mu$rad \cite{Scan1,Else1}.

Additionally, the distributions in Fig.~\ref{fig3} exhibit there is no the deflection at large angles ($\theta>14$ $\mu$rad) for very small miscut angles. It takes place due to very large terraces' lengths. Particles are deflected outward the surface by the single terrace (the type 1 of the motion) and have not the possibility to reach the planar channel. Finally, particles will be weakly reflected by the surface. The effective deflection of the beam at large angles through the particle quasichanneling (the type 3) in this case becomes  impossible.

Usually considered that the miscut brings only negative features to the crystal fabrication. In the letter we have demonstrated that the miscut surface, in principle, could be used to deflect the particle beam. It is important to underline that, on the contrary to the bent crystal technology, in the case of beam deflection by the miscut surface (we can define a new technique as a "miscut reflector") we deal with mostly reflection of the beam from the crystal surface; there is no necessity of using the crystal bulk to control the beam. Hence, the influence of the solid on the beam (scattering, energy loss, beam intensity loss, etc.) could be essentially reduced in comparison to both beam deflection by bent crystals and beam collimation by amorphous solids \cite{Scan7}. The problem to observe the phenomenon is in fabricating special crystals with controlled miscut angles. Nevertheless, the progress of crystal manufacturing technologies could issue the possibility to detect described peculiarities in the nearest future.

\end{document}